\documentclass[ reprint,onecolumn,12pt, nofootinbib]{revtex4-1}
\usepackage{graphicx}\usepackage{color}
\usepackage{amsmath}
\usepackage{graphicx}
\usepackage{dcolumn}
\usepackage{bm}
\usepackage{amsmath}
\usepackage{amsfonts}
\usepackage{color}

\def \lsim{\mathrel{\vcenter
     {\hbox{$<$}\nointerlineskip\hbox{$\sim$}}}}

\newcommand{\dd}{{\rm d}}
\newcommand{\beq}{\begin{equation}}
\newcommand{\eeq}{\end{equation}}
\newcommand{\beqa}{\begin{eqnarray}}
\newcommand{\eeqa}{\end{eqnarray}}
\newcommand{\bea}{\begin{eqnarray}}
\newcommand{\eea}{\end{eqnarray}}
\newcommand{\beqar}{\begin{eqnarray*}}
\newcommand{\eeqar}{\end{eqnarray*}}

\begin{tiny}

\end{tiny} 

\begin{document}

\title{Production of unstable heavy neutrinos in proto-neutron stars}

\author{C.~Albertus\footnote{conrado.albertus@dfa.uhu.es}, 
M.~Masip\footnote{masip@ugr.es}, 
M. A.~P\'erez-Garc\'ia\footnote{mperezga@usal.es}}

\affiliation{$^a$Departamento de F{\'\i}sica Aplicada, Universidad de Huelva, E-21071 Spain\\$^b$CAFPE and Departamento de F{\'\i}sica Te\'orica y del Cosmos,
Universidad de Granada, E-18071 Spain\\$^c$Department of Fundamental Physics, University of Salamanca, Plaza de la Merced s/n E-37008 Spain  }

\date{\today}
\begin{abstract}
We discuss the production of a class of heavy sterile neutrinos 
$\nu_h$ in proto-neutron stars. The neutrinos, of mass 
around $50$ MeV, have a negligible mixing with the active species but  
relatively large dimension-5 electromagnetic couplings.
In particular, a magnetic dipole moment $\mu\approx 10^{-6}$ GeV$^{-1}$ 
implies that they are thermally produced through $e^+ e^-\to \bar \nu_h \nu_h$
in the early phase of the core collapse, whereas a heavy--light 
transition moment $\mu_{\rm tr}\approx 10^{-8}$ GeV$^{-1}$ 
allows their decay $\nu_h\to \nu_i \gamma$ with a lifetime around $10^{-3}$ s.
This type of electromagnetic couplings has been recently proposed 
to explain the excess of electron-like events in baseline experiments. 
We show that the production and decay of these heavy neutrinos 
would transport energy from the central regions 
of the star to distances $d\approx 400$ km, providing a very 
efficient mechanism 
to enhance the supernova shock front and heat the material behind it. 
\end{abstract}

\maketitle

\section{Introduction}

Neutrinos define a sector of the Standard Model that  
still presents some important unknowns. The current scheme of mass differences
and mixings seems able to explain most of the existing data \cite{oscillation},
but the absolute value of their masses, their Dirac or Majorana 
nature \cite{maj} or the presence of additional sterile 
modes \cite{valle, Kusenko:2009up} are yet to be determined. In particular, 
the production of sterile neutrinos $\nu_s$, through collisions with 
standard matter or flavor oscillations has 
important implications in particle physics \cite{hidaka, wu, silkplb, daigne}. The mixing with an active neutrino $\nu$ may provide sterile modes 
with small couplings to the $W$ and $Z$ gauge bosons that translate into  
dimension-6 operators of type
\beq
-{\cal L}_{\rm eff}=
{G_F \sin \theta\over \sqrt{2}}\, 
\bar f \gamma_\mu ( C_V - C_A \gamma_5) f \, 
\bar \nu_s \gamma^\mu ( 1 - \gamma_5 ) \nu + {\rm h.c.}
\label{eqaa}
\eeq
In addition, the low-energy effective Lagrangian may also include 
dimension-5 operators from loops involving heavy particles.
%
Although these 
operators are usually overlooked, they could mediate the dominant 
reactions of sterile neutrinos in a star under favorable thermodynamical 
conditions. Here we will study this possibility in the context
of supernova explosions. 

When a supernova goes off a proto-neutron star can be formed having 
typical initial radius (20--60) km and (1--1.5)$M_\odot$ mass. It is 
believed that most of the gravitational binding energy 
($E_{\rm grav} \approx 3\times 10^{53}$ erg) is released in a $\sim$20 
second neutrino burst \cite{book}. The neutrino spectrum from   
supernova SN1987A detected at SuperK and IMB  
indicated a weak decoupling from baryonic matter, confirming 
that neutrino transparency sets in as their temperature falls 
below a few MeV \cite{koshiba} in the dense core.
At earlier phases of the collapse, however, 
computational simulations \cite{Pons:1998mm, fischer}
reveal internal peak 
temperatures exceeding 20 MeV in the central high density regions 
of the star. At such temperatures and densities 
the evolution of these astrophysical objects becomes sensitive 
to the fundamental properties of neutrinos and to the presence of
hypothetical weakly-coupled particles. 
In this context, a lot of effort has been devoted to the cooling 
through neutrino emission in the nuclear 
medium \cite{epja_perez}, to the matter opacity \cite{horowitz_05} 
and revival of the stalled shock that arises in the standard paradigm 
of supernova core collapse \cite{marek}, or to the synthesis of 
heavy nuclei taking place in the hot bubble behind the 
shock \cite{Janka:2006fh,Raffelt:2012kt}.

In this work we will focus on the astrophysical consequences of the production of a 
heavy sterile neutrino $\nu_h$ whose dominant interactions are {\it not}
the weak ones in Eq.~(\ref{eqaa}) but of electromagnetic kind. This type of particles
have been proposed as a possible explanation \cite{manel} 
for the excess of electron-like
events in baseline experiments \cite{miniboone}. Let us briefly show how
the required couplings could be generated.
Consider a $SU(2)_L$-singlet Dirac neutrino, $\nu_h$, of mass 
$m_h=50$ MeV. We will denote by $N$ and $N^c$ the (2-component) neutrino 
and antineutrino spinors defining $\nu_h$,
\beq
\nu_h=
\begin{pmatrix}N \cr \bar N^c 
\end{pmatrix}.
\eeq
Let us also suppose that at TeV energies the gauge symmetry is
$SU(2)_L\times SU(2)_R\times U(1)_{B-L}$ and that 
$\nu_{h}$ is accommodated within two $SU(2)_R$ 
doublets together with a charged lepton,
\beq
L=\begin{pmatrix}N\\E\end{pmatrix}\,\,\,\,
L^c=\begin{pmatrix}E^c\\N^c\end{pmatrix}.
\eeq
In order to avoid collider bounds \cite{bounds}, the breaking of 
the left-right symmetry must be such that
the charged lepton $(E,E^c)$ gets a mass $m_E\ge 300$ GeV while $\nu_h$ remains 
light. Loop diagrams of heavy gauge bosons and fermions (see Fig.~1) 
will then generate the operator
\begin{figure}
\begin{center}
\includegraphics[scale=0.5]{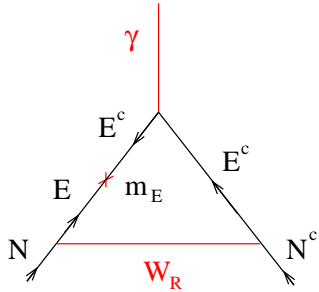}
\end{center}
\label{eenunu}
\vspace{-0.3cm}
\caption{Diagram contributing to the magnetic dipole moment $\mu$ of $\nu_h$.}
\end{figure}
\beq
-{\cal L}_{\rm eff}=
\mu\, \bar \nu_h \sigma _{\mu \nu} \nu_h\, 
\partial^\mu A^\nu \,,
\label{lef1}
\eeq
where $A^\nu$ is the electromagnetic field and $\mu$ is a 
magnetic dipole moment of order \cite{Bueno:2013mfa}
\beq
\mu \approx e\, {g_R^2\over 16
 \pi^2} \, {m_E\over M_R^2} \approx 10^{-6}\,{\rm GeV}^{-1}\,.
\eeq
In addition, the possible mixing of the sterile and the 
active neutrinos will be parametrized by an angle $\theta$, 
so that the mass eigenstates read $N'=\cos\theta\, N + \sin\theta \, \nu$ and 
$\nu'=-\sin \theta \, N + \cos \theta \,\nu$. 
This mixing will generate electromagnetic transitions 
through the same type of diagrams (we drop the prime to indicate
mass eigenstates):
\beq
{\cal L}_{\rm eff}=
{1\over 2} \, \mu_{\rm tr} \,
\overline \nu_h\, \sigma _{\mu \nu}
 \left(1-\gamma_5\right) \nu  \, \partial^\mu A^\nu + {\rm h.c.}\,, 
\label{int}
\eeq
with $\mu_{\rm tr}\approx \sin\theta \mu$ being the transition dipole moment. 
This operator may imply that the dominant decay mode of the heavy 
neutrino is $\nu_h \to \nu\,\gamma$.
Notice also that the presence of additional  
heavy singlets ($\nu_{h'}$ with $m_{h'}\approx m_E$) mixed
both with $\nu$ and $\nu_h$ will give additional contributions to 
$\mu_{\rm tr}$. Therefore, at this point we will treat $\mu$ 
and $\mu_{\rm tr}$ as independent parameters.

This type of sterile neutrinos could change substantially the 
evolution of a proto-neutron star. We will show that sterile pairs 
can be produced abundantly during the $\sim$20 second neutrino burst, 
escape the star core more easily than standard neutrinos, and finally 
decay within a few hundred km from the core.  The very energetic 
photons from the decay could deposit energy, helping revive the stalled 
accretion shock formed during the collapse
and change the thermal environment in the vicinity of the star. 
Our scenario could be considered a different realization of the 
{\it eosphoric} neutrino hypothesis proposed in \cite{Fuller:2009zz}.

\section{Decay rate, production and scattering cross sections}

Let us first describe the dominant decay and production channels 
for the heavy neutrino $\nu_h$ in vacuum. Later we will discuss how the 
hot and dense medium in a proto-neutron star (including populations of 
neutrons, protons, electrons and muons) affects these processes. 

To be definite in our calculation we will take as reference 
values $m_h = 50$ MeV, $\mu=10^{-6}\; {\rm GeV}^{-1} = 3.3\times
10^{-9}\mu_B$, and $\mu_{\rm tr}\approx 10^{-8}$ GeV$^{-1}$. We use $\hbar=c=1$. 
For these values of the mass and the transition moment, 
the heavy neutrino will decay into $\nu\gamma$  
with a lifetime
\beq
\tau_h =
{16 \pi \over \mu^2_{\rm tr}\, m_h^3}=
{(50\; {\rm MeV})^3 \over m_h^3}\times
{(10^{-8}\; {\rm GeV}^{-1})^2\over \mu_{\rm tr}^2}
\times 0.0026 \;{\rm s}\,.
\label{lifetime}
\eeq
We will also 
assume that the mass mixing, {\it i.e.}, the active component in $\nu_h$,
is smaller than $\sin^2\theta<10^{-3}$ and only along the muon and/or
the tau flavors. 
In that case, the radiative decay will dominate over the weak
processes $\nu_h\to \nu e^+ e^-\,,\, \nu \nu_i {\bar \nu_i}$, which appear
with a branching ratio
\beq
{\rm BR}(\nu_h\to \nu e^+ e^-)\approx 
{\sin^2\theta \over 10^{-3}}\times
{(10^{-8}\; {\rm GeV}^{-1})^2\over \mu_{\rm tr}^2}\times
{m_h^2 \over (50\; {\rm MeV})^2} \times 0.05\% \,.
\label{lifetime}
\eeq
This type of sterile neutrino avoids cosmological bounds since it 
decays before primordial nucleosynthesis. At colliders it is
hardly detectable: 
even if it were produced in $1\%$ of 
kaon or muon decays, 
$\nu_h$ is too long lived to
decay inside the detectors and too light to change significantly
the kinematics of the decay \cite{Gninenko:2010pr}. Actually, 
more elaborate setups with
two sterile modes have been proposed to explain the excess
of electron-like events at MiniBooNE in terms of the photon 
that results from its decay \cite{manel}.

\begin{figure}
\begin{center}
\includegraphics[width=8.cm]{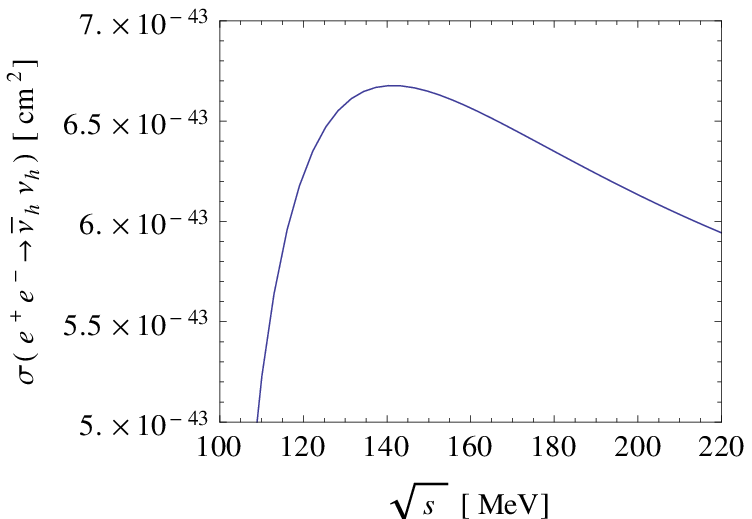}\hspace{0.3cm}
\includegraphics[width=8.cm]{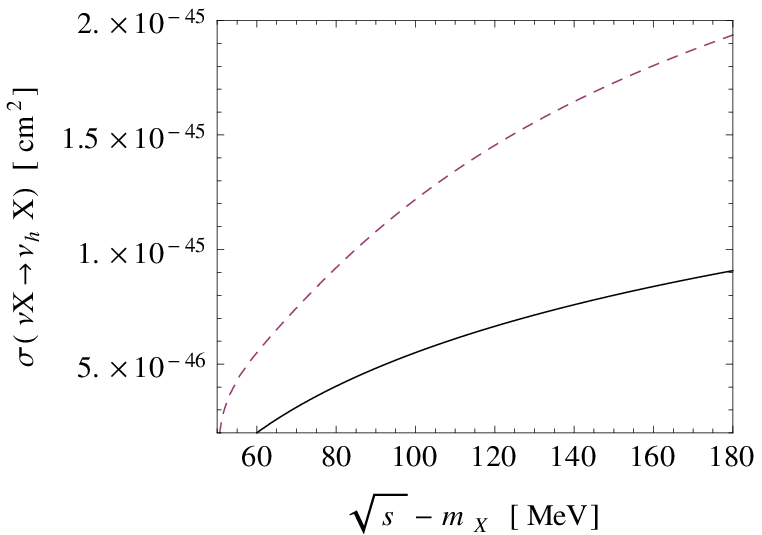}
\end{center}
\label{eenunu}
\vspace{-0.3cm}
\caption{
Left: Total cross section $\sigma(e^+ e^- \to \bar \nu_h \nu_h)$ 
for $m_h=50$ MeV and $\mu= 10^{-6}$ GeV$^{-1}$. Right: 
$\sigma(\nu X \to \nu_h X)$ 
for $\mu_{\rm tr}= 10^{-8}$ GeV$^{-1}$ 
(solid: $X=p$, dashes: $X=e$). 
}
\end{figure}

The dominant production channels of $\nu_h$ will also be 
electromagnetic. In particular, 
electron-positron annihilation  into $\nu_h$ pairs,
$e^+ e^- \to \bar \nu_h \nu_h$, will be  mediated 
by a photon through the magnetic dipole moment coupling in Eq.~(\ref{lef1}). The 
differential cross section is given by
\beq
{{\rm d}\sigma \over {\rm d} t}=
{\alpha \mu^2\over s^2-4sm_e^2}\left(
-t+2m_h^2+m_e^2-{t^2-2(m_h^2+m_e^2) t+(m_h^2-m_e^2)^2\over s}\right)
\,,
\eeq
where $\alpha$ is the fine structure constant, $m_e$ is the electron mass, 
$s$ and $t$ are the usual Mandelstam variables 
($\sqrt{s}$ is the center-of-mass energy). 
In Fig.~2 (left panel) we plot the total cross section for this process. Muon pair annihilation will give an analogous but subleading contribution,
since muons are less abundant than electrons in the star core.

 The active to sterile transition mediated by a photon
can be catalyzed by the presence of charged particles $X=p,e$: 
$\nu X \to \nu_h X$ (see right panel in Fig.~2). This contribution, 
however, can be neglected here due to the smaller value of the 
transition coupling that we have assumed, $\mu_{\rm tr}\approx 10^{-2}\mu$.
The weak channels which dominate the production of active neutrinos
\cite{buras} give also a subleading contribution due to the small mixing 
$\sin^2\theta < 10^{-3}$ of our sterile, whereas other
processes like plasmon decay \cite{plasmon} are irrelevant for heavy neutrino
masses around $50$ MeV ( {\it i.e.}, much larger than the electron mass).

In addition to its production and decay, the collisions of $\nu_h$ 
with charged matter will be essential in order to understand its propagation 
in the dense medium 
and estimate how efficiently these neutrinos escape the proto-neutron star. 
We need to distinguish between elastic scatterings 
\beq
\nu_h\, X \to \nu_h\, X
\eeq
and absorption reactions of type 
\beq
\nu_h\, X \to \nu\, X\,.
\eeq
The differential cross section for the first process reads
\beqa
{{\rm d}\sigma \over {\rm d} t}&=&
{\alpha \mu^2\over s^2-2s\left(m_X^2+m_h^2\right)+\left(m_h^2-m_X^2\right)^2}
\times \nonumber \\
&&\left(
-s+2m_h^2+m_X^2-{s^2-2(m_h^2+m_X^2) s+(m_h^2-m_X^2)^2\over t}\right)
\,.
\eeqa
This is a long distance (photon-mediated)
process with a divergent total cross section; if we restrict to
collisions substantially changing the direction of the incident
$\nu_h$ ({ {\it e.g.}}, a scattering angle $\theta> 30^o$ in 
the center-of-mass 
frame) we obtain the cross section depicted in Fig.~3 (left panel). 
\begin{figure}
\begin{center}
\includegraphics[width=7.8cm]{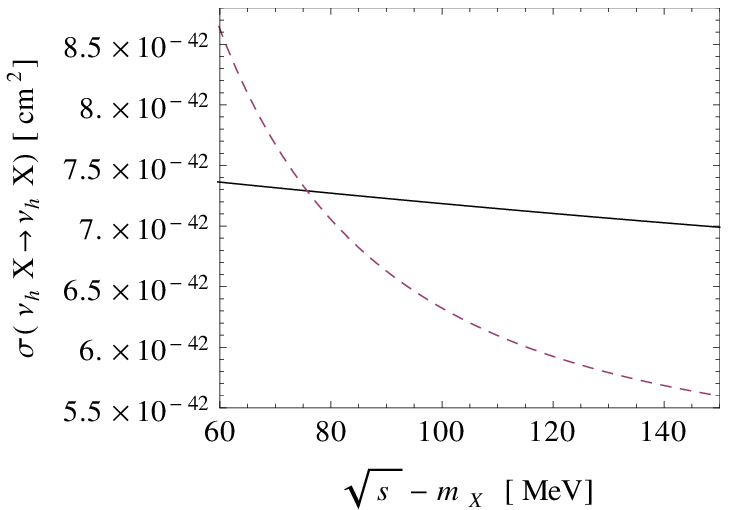}\hspace{0.5cm}
\includegraphics[width=7.8cm]{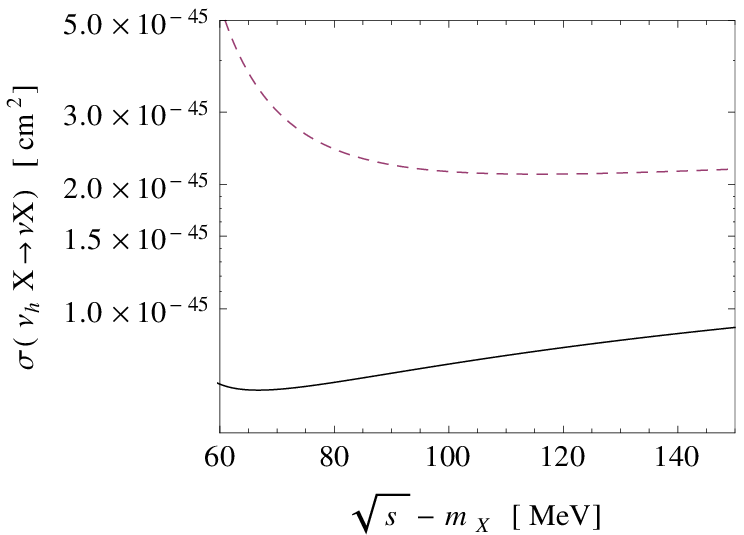}
\end{center}
\label{eenunu}
\vspace{-0.3cm}
\caption{
Left: $\sigma(\nu_h X \to \nu_h X)$ for 
$m_h=50$ MeV, $\mu= 10^{-6}$ GeV$^{-1}$ and a 
scattering angle $\theta>30^o$ in the center-of-mass frame. 
Right: $\sigma(\nu_h X \to \nu X)$ for 
$\mu_{\rm tr}= 10^{-8}$ GeV$^{-1}$ (solid: $X=p$, dashes: $X=e$)
}
\end{figure}
For the inelastic process that transforms the heavy neutrino
into an active one we obtain 
\beqa
{{\rm d}\sigma \over {\rm d} t}&=&
{\alpha \mu_{\rm tr}^2  \over 
2 \left( s^2-2s\left( m_X^2+m_h^2\right) +\left(m_h^2-m_X^2\right)^2\right)} 
\times \nonumber \\
&&\left(
-s+{1\over 2} (m_h^2+2m_X^2) 
  -{s^2-(m_h^2+2m_X^2) s+ {1\over 2} m_h^4 + m_X^4\over t}
-{m_X^2 m_h^4\over t^2}
\right)
\,,
\eeqa
and its  total cross section is also depicted in Fig.~3 (right panel). 
These cross sections are, in both cases, much smaller than the 
ones for active neutrinos off nucleons mediated by weak bosons.
For example, a neutrino in a gas at $T\simeq 20$ MeV has an 
average energy $\langle E_\nu \rangle \simeq \pi T \approx 60$ MeV, 
and its lowest order elastic cross section with a neutron 
is $\sigma(\nu_i n \to \nu_i n) \approx G^2_F E_\nu^2 (3 C^2_A+C^2_V)/\pi 
\simeq 3 \times 10^{-40}$ cm$^2$, with $i=e, \mu,\tau$ 
(the cross section with a proton is 
approximately hundred times smaller). As for the absorption of 
a $\nu_e$ through a charged current interaction, we have 
$\sigma(\nu_e n \to e p)\approx 2\times 10^{-39}$ cm$^2$.

\section{Production in a proto-neutron star}

Let us now calculate the production rate of heavy neutrinos
at the astrophysical site. Nucleon and lepton densities in the medium are constrained by electric charge neutrality and baryon
and lepton number conservation. Typical 
baryonic densities in the core of a proto-neutron star 
are well above nuclear saturation density $n_B\simeq$ (2--3)$n_0$ 
with $n_0=0.17$ $\rm fm^{-3}$ \cite{book}, whereas the lepton 
and electron fraction evolve dynamically from 
$Y_L\approx 0.31$, $Y_e\approx 0.27$ at $t=0.1$ s to 
$Y_L\sim0.18$, $Y_e\sim 0.17$ at $t=10$ s \cite{fischer_private}.
The extreme conditions in the core are such that quantum 
effects will be important. The population of baryons and leptons
is described by the Fermi-Dirac distribution
\beq
f_i(E)={1\over e^{(E-\mu_i)/T}+1}\,,
\eeq
where $\mu_{i}$ ($i=n, p, e^{\pm}$) denotes the chemical potential of the 
 considered species. Both $T$ and $\mu_{i}$
evolve within the star, in particular, 
the chemical potentials take care of the conservation of 
charges and quantum numbers in a self consistent way \cite{weak}. 

For a given value of the temperature and the electron chemical 
potential ($\mu_{e^+}=-\mu_{e^-}$), 
the {\it total energy emissivity} (energy produced per unit volume 
and unit time) of heavy neutrino pairs through the dominant process 
$e^+ e^-\to \bar \nu_h \nu_h$, $Q_E (e^+ e^-\to \bar \nu_h \nu_h) = \frac{\dd E}{\dd t \dd V}$ is given by \cite{yueh, Misiaszek:2005ax}
\small
\begin{equation}
Q_E=\frac{4}{(2 \pi)^8} \int \frac{\dd^3p_1}{2E_1} \frac{\dd^3p_2}{2E_2} 
\frac{\dd^3p_3}{2E_3} \frac{\dd^3p_4}{2E_4}\, \delta^4 (p_1+p_2-p_3-p_4) \,
(E_1+E_2)\, |\bar {\cal M}|^2\, f(f_1,f_2,f_3,f_4)
\label{qe}
\end{equation}
\normalsize
where the factor $f(f_1,f_2,f_3,f_4)= f_1 f_2 (1-f_3)(1-f_4)$ includes the 
Pauli blocking factor in the generic reaction $12\rightarrow 34$ and  
$p_i=(E_i, \vec p_i)$ are
the 4-momenta. We will consider that the reaction is 
not affected by the quenching of outgoing sterile states, {\it i.e.}, 
$(1-f_3) \simeq 1 \simeq (1-f_4)$, since heavy neutrinos
do not achieve chemical equilibrium and their number density 
inside the  star is always small. 
The squared matrix element for the interaction defined in 
Eq.~(\ref{lef1}) is given by 
\beq
|\bar {\cal M}(e^+ e^- \to \bar \nu_h \nu)|^2 = 
4e^2 \mu_h^2 \left(
-t+2m_h^2+m_e^2-{t^2-2(m_h^2+m_e^2) t+(m_h^2-m_e^2)^2\over s}\right)
\eeq
with $e$ the electron charge, $s=(p_1+p_2)^2$ and $t=(p_1-p_3)^2$,  see Fig.~4.
\begin{figure}
\begin{center}
\includegraphics[width=8.cm]{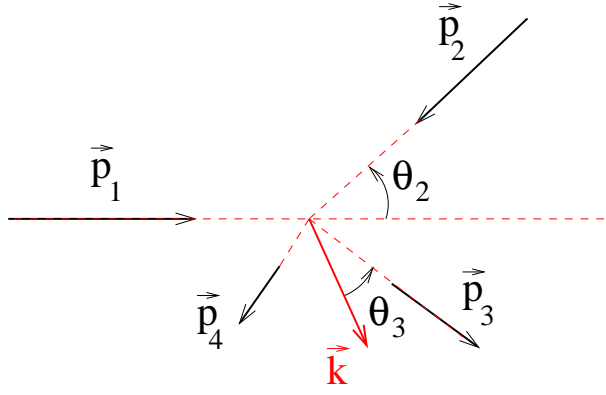}
\end{center}
\label{kinfig}
\vspace{-0.3cm}
\caption{Kinematical variables in the reaction $12\to 34$ used in this work.
}
\end{figure}

To perform the phase space integral in Eq.~(\ref{qe}) we  
note that there are only four non-trivial
independent variables: the initial energies 
$E_1$ and $E_2$, the angle 
$\theta_2$, as defined in Fig.~4, and $E_3$. Any other kinematical
variables can be derived from these four or can be trivially 
integrated. It is convenient to define  the 4-vector
$k=p_1+p_2$ (notice that $k^2=s$ and 
$|\vec k|^2= (E_1+E_2)^2-s$) and the angle $\theta_3$ of 
$\vec k$ with $\vec p_3$:
\beq
\cos \theta_3 = {2\, (E_1+E_2)\, E_3 -s\over 2\, |\vec k|\,
\sqrt{E_3^2 - m_h^2} }\,.
\eeq
After integrating the 4-dimensional Dirac delta
that enforces energy and momentum conservation, we obtain
\beq
Q_E=\frac{1}{64 \pi^5}\int_{m_e}^\infty \!\dd E_1 
\int_{E_2^{\rm min}}^\infty \!\dd E_2
\int_{c_{2}^{\rm min}}^1 \!\! \dd \cos\theta_2
\int_{E_-}^{E_+} \dd E_3\, {p_1 p_2 \over |\vec k|}
\left(E_1+E_2\right) |\bar {\cal M}|^2 \,f_{e^+}(E_1)\, 
f_{e^-}(E_2)\,,
\label{qeval}
\eeq
where the minimum values of $E_2$ and $\cos\theta_2$, 
$E_2^{\rm min}$ and $c_{2}^{\rm min}$, respectively,  result from the 
kinematical restriction $s>4m^2_h$ 
\beq
E_{2}^{\rm min}(E_1) \approx \sqrt{
\frac{m^4_h+m^2_e E^2_1}{ E^2_1-m^2_e } }\,, \,\,\,\,c_{2}^{\rm min}(E_1, E_2)
={\rm Max}\left[-1,\frac{2m^2_h-m^2_e-E_1 E_2}{p_1 p_2}\right] .
\eeq
We also define
\beq
E_{\pm}(E_1,E_2,\cos\theta_2) =
\frac{E_1+E_2}{2} \pm \frac{|\vec k|}{2} \sqrt{1-\frac{4 m^2_h}{s} }\,.
\eeq
The Mandelstam variables in terms of these four quantities read
\bea
s &=& 2 \left( m_e^2 + E_1 E_2 + p_1 p_2 \cos \theta_2 \right)\\
t &=& - 2 E_1 E_3 + 2 p_1 p_3 \cos (\theta_3 - \alpha)
+m_e^2 + m_h^2\, 
\eea
where $\alpha$ ($0\le \alpha \le \pi$) is the angle 
between $\vec k$ and $\vec p_1$,
\beq
\alpha = \arctan \left(
{p_2 \sin \theta_2\over p_1-p_2\cos\theta_2}\right) \,.
\eeq

\section{Transport of energy out of the star core}

The possible impact of the heavy neutrino $\nu_h$ on the 
evolution of the proto-neutron star will depend on its ability 
to take a significant amount of energy
out of the core. If the sterile neutrinos are abundant 
inside the star core but unable to reach the surface before decaying
into a photon plus an active neutrino, then they become just a state mediating 
interactions of electrons with neutrinos and photons.
We will show that this is not the case and that
they could play an interesting role in supernova explosions.

Let us take a temperature $T_0=25$ MeV and an electron 
chemical potential $\mu_{e0}=100$ MeV, which are typical values
at the inner central region of a proto-neutron star 
(see \cite{fischer,Pons:1998mm}). Although these quantities are 
time and density dependent, the chosen values can
be used to estimate the possibilities of our scenario.

Varying the mass $m_h$, the magnetic dipole moment $\mu$, the electron chemical potential $\mu_e$ and 
the temperature $T$ and performing a fit of $Q_E$ in Eq.(\ref{qeval}) 
we obtain 
\beq
Q_E\approx 1.5\times 10^{36} \left({\mu\over 10^{-6}\,{\rm GeV}^{-1}}\right)^2
\left({T\over 25\,{\rm MeV}}\right)^{7.4} e^{-{m_h +\mu_e \over 3 T}}
\; {\rm erg\over s\;cm^3}\,.
\eeq
For the reference values of all the parameters, the expression
above yields $Q_E\approx 2\times 10^{35}\; {\rm erg/ s\; cm^3}$, 
with an average $\nu_h$ energy of 
$\langle E_{h} \rangle \approx 103$ MeV. This is a very large production
rate, $\sim10^2$--$10^3$ times larger than the one obtained in \cite{Fuller:2009zz} using
heavier sterile neutrinos mixed with the active ones. Our neutrinos, however,
will not leave the star core unscattered.

We can also compare this production rate with the
one of standard neutrinos in early cooling of proto-neutron stars. 
For example, in the central core the direct URCA process $n\rightarrow p e {\bar \nu}_e$ provides 
$Q^{DURCA}_E\approx 2.4 \times 10^{41}R\; {\rm erg/ s\;cm^3}$ at $T=25$ MeV 
\cite{yueh,yueh2}, being $R$ a factor of order unity \cite{page}. 
This is five decades over
the $\nu_h$ production rate  that we have found. The direct 
URCA process requires a high proton fraction, $Y_p \gtrsim 11\%$, only accessible to large mass objects, however the less demanding modified URCA cooling also gives a much faster rate than for steriles,
$Q^{MURCA}_E\approx 1.5 \times 10^{40}R\; {\rm erg/ s\;cm^3}$. These
active neutrinos will be, to a large extent, trapped  (before transparency sets in) inside the
star core, whereas ours have weaker interactions with the protons and electrons in the medium.

It is then apparent that we need to
consider propagation effects of $\nu_h$. 
Although a precise calculation would 
require 
a complete multidimensional simulation that is beyond the scope of
this work, we will discuss the qualitative picture and show that the model
has enough parameter space to realize it.

The first important effect is the diffusion from the center to
the surface of the star core, with a radius $r\approx 20$ km.  
Active muon and tau neutrinos scatter there mainly off neutrons 
with a mean free path 
$\lambda_{\nu}^S\sim 1$ m \cite{Reddy:1997yr,Pons:1998mm}. 
Analogously, $\nu_h$ 
will scatter elastically with protons 
with a cross section$\sim$40 times smaller, which suggests
an interaction length $\lambda_{h}^S$ inside the star longer by the same factor. 
This implies a larger diffusion 
coefficient $D\approx \lambda_{h}^Sc/3$ and a more efficient
transport from the core to the outer parts of the star. 
The typical diffusion time for that process will be
$\tau_D \approx r^2/(2D)\approx 5 \times 10^{-2}$ s.
The second crucial effect in the propagation of the heavy neutrinos
is their absorption: the star will capture a fraction of them
through the inelastic collisions $\nu_h p\to \nu p$. 
For $\mu_{\rm tr}=10^{-8}$ GeV$^{-1}$ 
the absorption length is approximately $\sigma(\nu_hX\to \nu_h X)/
\sigma(\nu_h X \to \nu X) \approx 3000$ times
larger than $\lambda_{h}^S$, {\it i.e.}, 
$\lambda_{h}^A\approx 120$ km. A final effect, analogous to absorptions, is their decay.  
Since the values of $\mu_{\rm tr}$ and $m_h$ that we have assumed 
imply a sterile neutrino lifetime $\tau_h\approx 0.003$ s, 16
times smaller than their diffusion time, 
a large fraction of the heavy neutrinos produced in the core 
will decay into $\gamma \nu$ 
before they have diffused to the outer layers.  For a time window of $\sim20$ s energy can be transported to distances $d\simeq \sqrt{2 D \tau_D } \simeq 400$ km.

We estimate that, even if absorptions and decays reduced the number of 
neutrinos leaving the star from our estimate to $1\%$ of the ones produced (some of
them closer to the surface), $\nu_h$
may still carry a total of $10^{51}$--$10^{52}$ erg and 
deposit this energy outside the star during the 20 second neutrino burst. 
Notice that a reduction in the magnetic moment $\mu$ would also reduce
the production rate of heavy neutrinos, but it would increase the mean free
path between elastic scatterings and then the fraction of neutrinos
that reach the surface. In addition to the transition moment 
$\mu_{\rm tr}$, the mass $m_h$ is another parameter that could 
impact the production rate or the decay length of 
the heavy neutrino.

An interesting variation of the model that depends only on two
parameters ($m_h$ and $\mu_{\rm tr}$) would be obtained by 
suppressing the magnetic moment $\mu$ and slightly increasing 
the transition one, {\it e.g.} 
$\mu_{\rm tr}\approx 5\times 10^{-8}$ GeV$^{-1}$. In that case the dominant
production channel would be
\beq
\nu X \to \nu_h X\,;\hspace{0.5cm}\bar \nu X \to \bar \nu_h X\,,
\eeq
where $X$ is any charged particle and the active $\nu$ may be any 
linear combination of $\nu_\mu$ and $\nu_\tau$. In the 
star core the heavy neutrinos
would be partially absorbed through the inverse reaction (we estimate
an absorption length $\lambda^A_{h}\approx 4$ km), and the ones
reaching the surface would decay 
with $c\tau_h\approx 30$ km. Since the production would not be
so abundant as in the case discussed in the previous section, 
this possibility 
provides a {\it safer} scenario still able to transport energy to the
region near the star surface.

Although the optimal value of these parameters would require a 
full Monte Carlo simulation, the scenario seems
flexible enough to introduce acceptable changes in the dynamics of
supernova explosions, with the decay into photon plus
active neutrino playing an important role in the enhancement of
the supernova shock.

\section{Conclusions}

The collapse of a very massive star defines an astrophysical object
with extreme conditions where neutrinos determine the 
thermodynamics. These proto-neutron stars are suitable laboratories 
to probe the properties
of any weakly coupled particles of mass $\lsim 100$ MeV. Here we have
proposed a sterile neutrino $\nu_h$ much heavier than the standard ones
($m_h\approx 50$ MeV) 
and with sizable electromagnetic couplings: a magnetic dipole moment
$\mu$ and dipole sterile--active transition $\mu_{\rm tr}$ that mediates 
its decay $\nu_h\to \nu \gamma$ with a lifetime
$\tau_h\approx 10^{-3}$ s. This simple 3-parameter model seems to have 
interesting implications in supernova explosions. The heavy neutrino 
is produced
in the core at a high rate through $e^+ e^-\to \bar \nu_h \nu_h$, 
it may escape the star more efficiently than active neutrinos and decays depositing a large amount of energy in the outer layers of the star.

We believe that the type of heavy sterile neutrino proposed here 
could be an essential ingredient to help the progression of the 
internal shock, which is responsible for the observed supernova events. 
Full computational simulations could 
shed more light into the complex energy transport that results 
from competing processes of scattering, interaction and decay.

\section*{Acknowledgments}

We woud like to thank T.~Fischer and  J.~Pons for useful discussions.
This work has been supported by MICINN of Spain (FPA2013-47836, 
FIS2012-30926)  and  
Consolider-Ingenio {\bf Multidark} CSD2009-00064) and by Junta de 
Andaluc\'\i a (FQM225, FQM101,3048).


\end{document}